\documentclass[fleqn,usenatbib]{mnras}

\usepackage{newtxtext,newtxmath}

\usepackage[T1]{fontenc}

\DeclareRobustCommand{\VAN}[3]{#2}
\let\VANthebibliography\thebibliography
\def\thebibliography{\DeclareRobustCommand{\VAN}[3]{##3}\VANthebibliography}

\usepackage{graphicx}	
\usepackage{amsmath}	

\usepackage{graphicx}
\usepackage[caption=false]{subfig}
\usepackage{listings}
\usepackage{xparse}
\usepackage{xcolor}
\usepackage{dcolumn}
\usepackage{bm}
\usepackage{hyperref}

\usepackage{tabularx}

\newcommand\ege{Department of Astronomy and Space Sciences, Faculty of Science, Ege University, 35100, {\.I}zmir, Turkey}
\newcommand\egravity{Ege Gravitational Astrophysics Research Group (eGRAVITY), Ege University, 35100, {\.I}zmir, Turkey}
\newcommand\milano{Dipartimento di Fisica G. Occhialini, Universit\`a di Milano - Bicocca, Piazza dellaScienza 3, I-20126 Milano, Italy}
\newcommand\INFN{INFN, Sezione di Milano-Bicocca, Piazza della Scienza 3, I-20126 Milano, Italy}
\newcommand\INAF{INAF, Osservatorio Astronomico di Brera, Via E. Bianchi 46, I-23807 Merate, Italy}

\newcommand{\codeword}[1]{\texttt{#1}}
\newcommand{\model}[1]{\texttt{#1}}
\newcommand{\Mthres}{$M_{\rm thres}$}
\newcommand{\Msun}{$M_\odot\ $}

\newcommand{\M}{$M$}

\newcommand{\fmerger}{$f_{\rm merger}$}

\title[Simulations of High-Mass Binary Neutron Star]{General Relativistic Simulations of High-Mass Binary Neutron Star Mergers: rapid formation of low-mass stellar black holes}

\author[K. A. \c{C}okluk et al.]{
Kutay A. \c{C}okluk,$^{1,2}$\thanks{E-mail: kutay.arinc.cokluk@ege.edu.tr}
Kadri Yakut,$^{1,2}$
Bruno Giacomazzo$^{3,4,5}$
\\
$^{1}$\ege \\
$^{2}$\egravity \\
$^{3}$\milano \\
$^{4}$\INFN \\
$^{5}$\INAF \\
}

\date{Accepted 2023 December 1. Received 2023 November 19; in original form 2023 September 27}

\pubyear{2023}

\begin{document}
\label{firstpage}
\pagerange{\pageref{firstpage}--\pageref{lastpage}}
\maketitle

\begin{abstract}
Almost a hundred compact binary mergers have been detected via gravitational waves by the LIGO-Virgo-KAGRA collaboration in the past few years providing us with a significant amount of new information on black holes and neutron stars. In addition to observations, numerical simulations using newly developed modern codes in the field of gravitational wave physics will guide us to understand the nature of single and binary degenerate systems and highly energetic astrophysical processes. We here presented a set of new fully general relativistic hydrodynamic simulations of high-mass binary neutron star systems using the open-source \codeword{Einstein Toolkit} and \codeword{LORENE} codes. We considered systems with total baryonic masses ranging from $2.8$\Msun to $4.0$\Msun and used the SLy equation of state. We analyzed the gravitational wave signal for all models and reported potential indicators of systems undergoing rapid collapse into a black hole that could be observed by future detectors like the Einstein Telescope and the Cosmic Explorer. The properties of the post-merger black hole, the disk and ejecta masses, and their dependence on the binary parameters were also extracted. We also compared our numerical results with recent analytical fits presented in the literature and provided parameter-dependent semi-analytical relations between the total mass and mass ratio of the systems and the resulting black hole masses and spins, merger frequency, BH formation time, ejected mass, disk mass, and radiated gravitational wave energy.
\end{abstract}



\begin{table*}
\centering
\caption{We list the BNS systems studied in this paper. The first column lists the model names, e.g. \codeword{M32q10} refers to a BNS system with a total baryonic mass of $3.2 M_\odot$ and mass ratio $q=1.0$. While $M_{1b,2b}$ represent the baryonic mass of the components, $M_{1,2}$ are the gravitational mass of each neutron star at infinite separation. $M_0$, $J_0$ and $f_0$ are, respectively, the total initial ADM mass, the total initial angular momentum and the initial orbital frequency of the system. The last three columns show the compactness parameters of each NS and the combined dimensionless tidal deformability $\widetilde{\Lambda}$, all computed at infinite separation.}
\label{table:initialdata}
\begin{tabular}{lcccccccccr}
\hline
{Model}             &$M_{1b}$      &$M_{2b}$	    &$M_{1}$        &$M_{2}$        &$M_0$              &$J_0$	                                    &$f_0$                  &$C_{1}$    &$C_{2}$    &$\widetilde{\Lambda}$  \\
			        &$[M_\odot]$    &$[M_\odot]$ 	&$[M_\odot]$    &$[M_\odot]$    &$[M_\odot]$		&$\!\!\left[\frac{GM_\odot^2}{c}\right]$    &$\left[Hz\right]$                                                      \\
\hline
\codeword{M28q10}	&1.40			&1.40			&1.28           &1.28           &2.54				&6.41   	                                &324.64	                &0.161      &0.161      &511.90     \\
\hline
\codeword{M30q10}	&1.50			&1.50			&1.36           &1.36           &2.70				&7.11   	                                &332.41	                &0.172      &0.172      &343.70   	\\  
\codeword{M30q11}   &1.57			&1.43			&1.42           &1.30           &2.70				&7.09   	                                &332.38	                &0.179      &0.164      &347.14  	\\
\hline
\codeword{M32q10}	&1.60			&1.60			&1.44           &1.44           &2.85				&7.83   	                                &339.64	                &0.182      &0.182      &234.21  	\\  
\codeword{M32q11}	&1.68			&1.52			&1.50           &1.38           &2.85				&7.81  	                                    &339.49	                &0.191      &0.174      &236.73   	\\
\codeword{M32q12}	&1.75			&1.45			&1.56           &1.32           &2.85				&7.77   	                                &339.39	                &0.198      &0.167      &243.35 	\\
\codeword{M32q13}	&1.81			&1.39			&1.61           &1.27           &2.85				&7.70    	                                &339.23	                &0.206      &0.160      &253.33 	\\
\codeword{M32q14}	&1.87			&1.33			&1.65           &1.22           &2.85				&7.62                                       &339.11	                &0.212      &0.153      &265.62     \\
\codeword{M32q16}	&1.97			&1.13			&1.73           &1.14           &2.85				&7.42   	                                &338.64	                &0.224      &0.142      &297.33 	\\
\codeword{M32q18}	&2.06			&1.14			&1.79           &1.06           &2.85				&7.21   	                                &338.07	                &0.235      &0.133      &335.11  	\\
\codeword{M32q20}	&2.13			&1.07			&1.85           &1.00           &2.85				&6.98   	                                &337.76	                &0.244      &0.124      &377.94 	\\
\hline
\codeword{M34q10}	&1.70			&1.70			&1.52           &1.52           &3.01				&8.51   	                                &346.16	                &0.193      &0.193      &161.02  	\\  
\hline
\codeword{M36q10}	&1.80			&1.80			&1.60           &1.60           &3.16				&9.33   	                                &352.37	                &0.205      &0.205      &111.01  	\\  
\codeword{M36q11}	&1.89			&1.71			&1.66           &1.53           &3.16				&9.31   	                                &352.45	                &0.214      &0.195      &112.51  	\\
\hline
\codeword{M38q10}	&1.90			&1.90			&1.68           &1.68           &3.31				&10.11   	                                &358.42	                &0.216      &0.216      &76.46  	\\  
\codeword{M38q11}	&1.99			&1.81			&1.74           &1.61           &3.31				&10.09   	                                &358.25	                &0.226      &0.206      &77.61  	\\
\hline
\codeword{M40q10}	&2.00			&2.00			&1.75           &1.75           &3.46				&10.90   	                                &363.73	                &0.228      &0.228      &52.19  	\\  
\codeword{M40q11}	&2.10			&1.90			&1.82           &1.68           &3.46				&10.88   	                                &363.83	                &0.239      &0.216      &53.00  	\\
\hline
\end{tabular}
\end{table*}

\section{Introduction}
\label{sec:intro}

Since the first binary black hole merger observations detected by the LIGO-Virgo collaboration \citep{Abbott_2016}, a new window has been opened in astrophysics in the gravitational wave (GW) field \citep{Barack_2019}. In 2017, for the first time, the GW signal from a binary neutron star (BNS) merger, named GW170817, was also detected \citep{Abbott_2017a} and it was accompanied by electromagnetic (EM) counterpart observations from gamma rays to radio \citep{Abbott_2017b}. Thus, the multi-messenger era has begun. Two years after the first detection of a BNS merger, GW190425, a new BNS system, was detected by the LIGO-Virgo collaboration \citep{Abbott_2020}. For the latter, there is however no detected EM counterpart. GW190425 observation is significant due to being the heaviest BNS system ever observed, with a total mass ($3.4^{+0.3}_{-0.1} M_\odot$ assumption of high-spin prior) much larger than the one measured for the galactic BNS systems to date~\citep{Farrow_2019, Zhang_2020}.

The possible outcomes of a BNS merger can be either a prompt collapse to a black hole (BH), or the formation of a short-lived hypermassive neutron star (HMNS) or long-lived supermassive neutron star (SMNS) that eventually collapses to a BH, or a stable NS \citep{Piro2017}. If the total gravitational mass of the system, \M, is higher than a threshold mass, \Mthres (e.g., \citep{Hotokezaka_2011, Bauswein_2013, Koppel_2019,Barack_2019,Kashyap2022, Perego2022}), then the remnant of the merger will promptly collapse to a BH. Although from the galactic population of BNSs the masses of NSs in BNS systems are expected to lie in a range $1.3-1.4M_\odot$, studies \citep{Paschalidis_2019,Margalit_2019} suggest, respectively, that up to 25\% and 32\% of BNS mergers might produce a rapid collapse to BH. For GW190425, it is estimated that the probability of the binary promptly collapsing into a BH after the merger is 96\% , with the low-spin prior, or 97\% with the high-spin prior \citep{Abbott_2020}.
 
BNS simulations in recent years (see, e.g., \citet{Shibata_2003, Shibata_2005, Shibata_2006, Kiuchi_2009, Rezzolla2010, Hotokezaka_2013, Kastaun2013, East_2016, Endrizzi_2016, Dietrich_2017a, Dietrich_2017b, Endrizzi2018, Paschalidis_2019,  Most_2019, East_2019, Ruiz_2019,  Ruiz_2020, Tootle2021, Papenfort2022, Sun_2022})  have revealed that the amount of disk mass surrounding BH and the amount of ejected matter from the system are strongly dependent on the mass of the system, mass ratio, equation of state (EOS), spin, and magnetic field of neutron stars. In addition, the effects of the BNS mass, mass ratio, EOS, spin, and magnetic field constitute a highly degenerate parameter space, which makes precise arguments difficult to make, as long as simulations that include all of the above are still lacking. Because nearly all matter is swallowed by the remnant BH, it is expected that equal-mass BNS systems with \M$>$\Mthres\ will have a disk with negligible mass and a low amount of ejected matter.

High-mass mergers, such as the ones discussed in this study, are important to study because they can provide insight into high-mass neutron stars and the remnants produced by their mergers. We can also learn about the internal structure of massive neutron stars (e.g., their EOS) by analyzing these types of mergers. We conducted simulations of high-mass binary systems that underwent rapid collapse in this study, and we investigated the effects of mass ratio and total mass on gravitational wave emission and system dynamics. The paper is organized as follows: we review the numerical setup and initial data for our models in Section~\ref{sec:setup}. The numerical results are presented in Section~\ref{sec:results}. Finally, we discuss our findings in Section~\ref{sec:conclusions}. We use a system of units in which $c=G=M_\odot=1$, unless specified otherwise.


%
\begin{figure*} 
\centering
\subfloat{{\includegraphics[width=.45\linewidth]{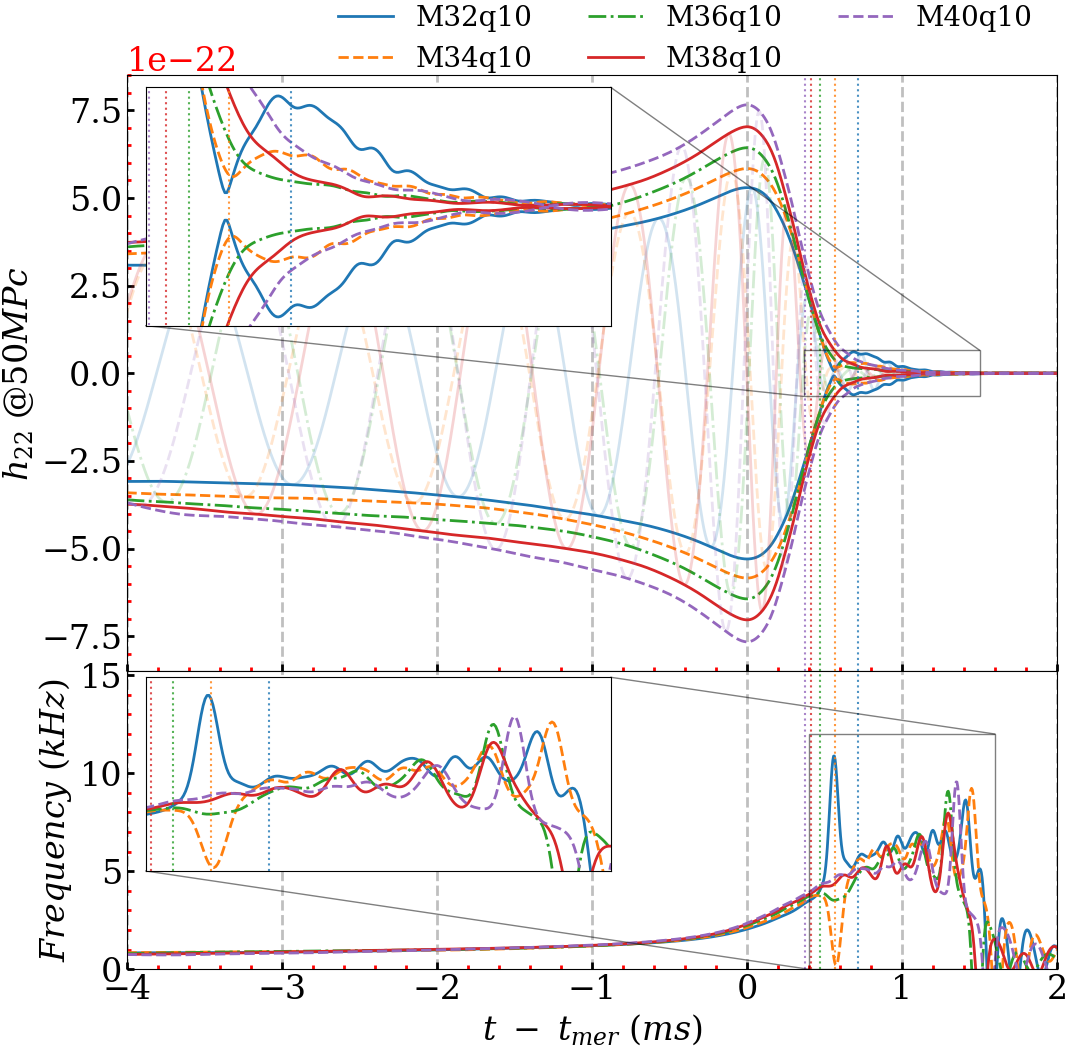} }}
\qquad
\subfloat{{\includegraphics[width=.45\linewidth]{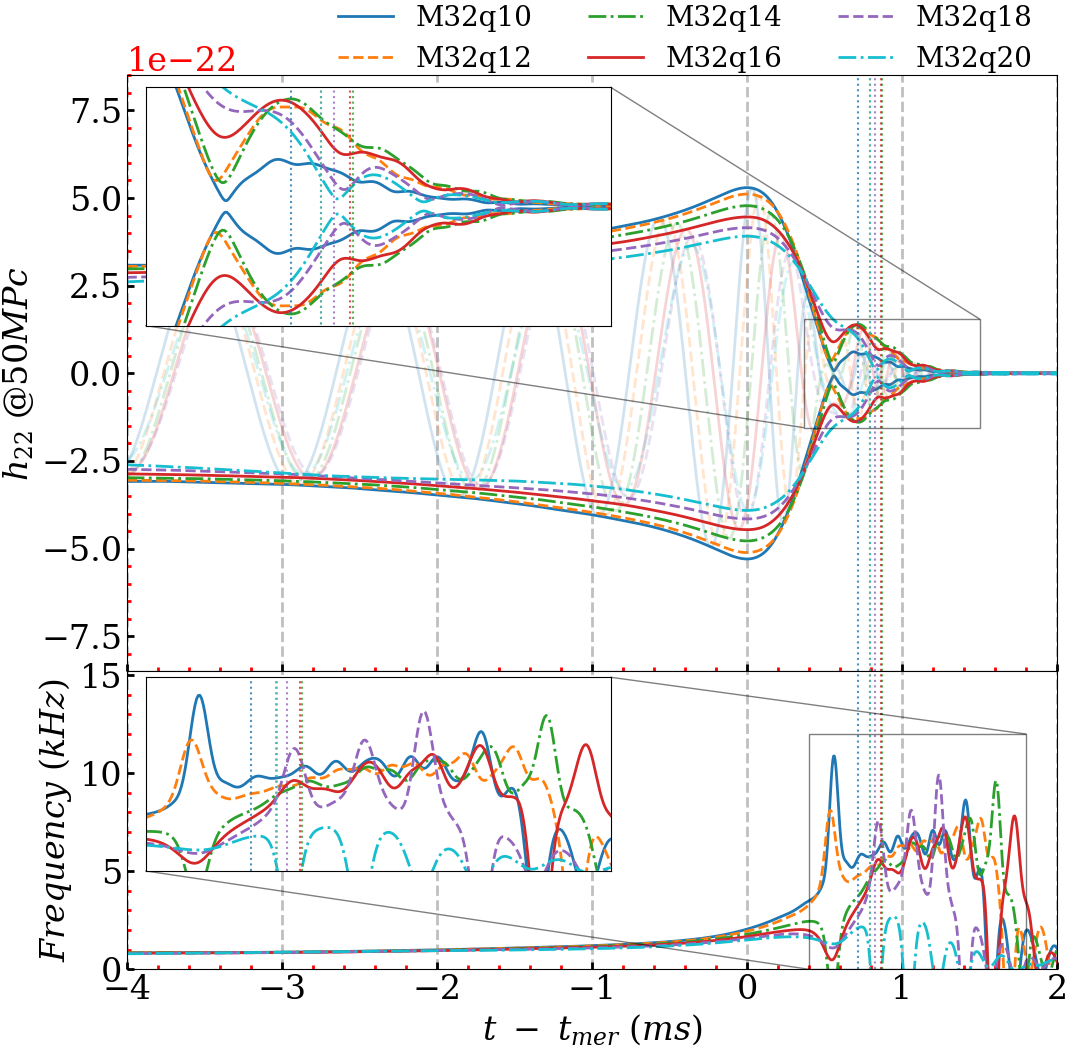} }}
\caption{\label{fig:GWFreq} The left and right figures represent, respectively, equal and unequal mass models. Upper panels: extracted gravitational waveforms of the dominant mode (l,m=2,2) computed at $r=444km$. The thin lines show the GW signal $h_{22}=h_{+,22}$, while the thick outer lines represent the GW amplitudes, $|h|=\sqrt{|h_+|^2 + |h_\times|^2}$ and it does not take into account the effect of the viewing angle. Bottom panels: instantaneous frequency, i.e. the time derivative of the GW phase of each model. Note that the instantaneous frequency is smoothed by convoluting with a blackman window function in $0.1\ ms$.}
\end{figure*}
%

\section{Models and Numerical Setups}
\label{sec:setup}

We consider a set of irrotational equal and unequal mass BNS systems with a total baryonic mass between $2.8-4.0$\Msun. We computed the initial data using the pseudo spectral elliptic solver \codeword{LORENE}~\citep{Gourgoulhon_2001, LORENEcode} assuming irrotational NSs on a quasi-circular orbit. For all models, the initial coordinate separation is $40~km$, corresponding to $\approx 3$ orbits before the merger. We report the initial parameters of our models in Table \ref{table:initialdata}. The first column refers to the names of the models, e.g. \model{M32q10} means that the initial total baryonic mass of the BNS is $3.2$\Msun and that the mass ratio is $1.0$ (the mass ratio is defined such that $q = M_{1b}/M_{2b} > 1$). From the second to fifth columns we show, respectively, neutron stars' initial baryonic masses ($M_{1b,2b}$) and gravitational ($M_{1,2}$) masses at infinite separation. $M_0$, $J_0$ and $f_0$ are the values of the initial ADM mass, total angular momentum, and orbital frequency. The compactness parameters of each NS are given as $C_{1,2}$ where $C_{1,2} = GM_{1,2}/ R_{1,2}c^2$ and $C_1$ and $C_2$ are computed considering the NS at infinite separation. The last column refers to the reduced tidal parameter, $\widetilde{\Lambda}$ \citep{Favata_2014}: 
\begin{eqnarray}
\widetilde{\Lambda} =  \frac{16}{13}\frac{(M_1 + 12 M_2)}{M^5} \Lambda_1 {M_1^4}+ (1 \rightarrow 2)\,,
\label{eq:one}
\end{eqnarray} 
where the quadrupolar tidal parameter of the individual stars \citep{Flanagan_2008,Damour_2010} is $\Lambda_i = \frac{2}{3} C_i^{-5} k_i^{(2)}, i=1,2$, and $k_i^{(2)}$ is the dimensionless Love number \citep{Damour_1983,Hinderer_2008,Damour_2009,Binnington_2009}. The notation $(1 \rightarrow 2)$ indicates a second term identical to the first, but 1 and 2 are exchanged.

This work employs the SLy EOS \citep{Douchin_2001} to describe the NS matter. To build the initial data, we used the EOS table provided by Parma Gravity Group\footnote{https://bitbucket.org/GravityPR}. During the evolution, we instead used the seven piece polytropic version (SLyPP) described in \citet{DePietri_2016} and added a thermal component given by $\Gamma_{\rm thermal} = 1.8$.

To solve the general relativistic hydrodynamic (GRHD) equations, we used the publicly available \codeword{Einstein Toolkit} (ET) \citep{ETKcode, Loffler_2012} code, based on \codeword{Cactus Computational Toolkit}\footnote{http://www.cactuscode.org/}. In particular, we used the nineteenth release version of ET named ``Katherine Johnson'' (ET\_2021\_11) \citep{2021zndo...5770803B}. 

We evolve the space-time metric using the BSSN formalism \citep{Nakamura_1987,Shibata_1995,Baumgarte_1998,Alcubierre_2000,Alcubierre_2002} as implemented in the \codeword{McLachlan} thorn\footnote{http://www.cct.lsu.edu/~eschnett/McLachlan}. We use the \codeword{GRHydro} code \citep{Baiotti_2004,Hawke_2005,Moesta_2013} to solve the GRHD equations and employ a fifth-order WENO-Z reconstruction method \citep{Borges_2008}. The time integration is handled with a fourth-order Runge-Kutta method \citep{Runge_1895, Kutta_1901} with a Courant factor of $0.4$. We also set an artificial atmosphere of $\rho_{atm} = 1.01 \times 10^{-11}$. We also would like to state that the value of $\rho_{atm}/\rho_{max}$ for the less massive companion, which is from model \model{M32q20} and has $1.07 M_\odot$ baryonic mass, is $9.08 \times 10^{-9}$.

We employ an adaptive mesh refinement (AMR) approach provided by the \codeword{Carpet}  driver \citep{Carpetcode} and we force the finest grids to follow each NS during the inspiral. The grid hierarchy consists of seven nested refinement levels with a $2:1$ refinement factor. The resolution of each simulation is characterised by the resolution of the innermost grid, which is $dx = 0.16 \approx 237$ m, while the radius of the outer boundary of the numerical domain is $1024 \approx 1514$ km. We apply reflection symmetry along the z=0 plane. After merger, if the final remnant's mass exceeds the EOS's threshold mass, the \codeword{AHFinderDirect} thorn \citep{Thornburg_2003} is used to detect the formation of an apparent horizon and extract the properties of the BH.

  Note that results presented in this paper are extracted from the simulations having the standard resolution (SR) of $dx=0.16$. For two models, we also run a high resolution (HR) and low resolution (LR) simulations with $dx=0.12$ and $dx=0.20$ to estimate our numerical error. The accuracy of our results is discussed in Appendix \ref{appendix:A}.


\section{Results}
\label{sec:results}

\subsection{GW Extraction and Analysis}
\label{sec:gw}
We computed the GWs from $-7\ ms$ to $10\ ms$, where $t=0\ ms$ corresponds to the maximum strain amplitude and refers to the merger time, and the GWs include approximately the last three orbits before merger. 

We extract the GW signal at a coordinate radius of $\approx 444\ km$ from the BNS center of mass, calculating the Newman-Penrose scalar $\psi_4$ \citep{Newman_1962} (Eq. \ref{eq:psi4}), provided by ET module \codeword{WeylScal4}
, and decomposed in spin-weighted spherical harmonics using the ET module \codeword{Multipole}:
\begin{equation}
\ddot{h}_+ - \ddot{h}_\times = \psi_4= 
\sum_{l=2}^{\infty} {\sum_{m=-l}^{l} {\psi_4^{lm}(t,r)\ {{}_{-2}\!}{Y}_{lm}(\theta,\phi)}}.
\label{eq:psi4}
\end{equation}
Since $\psi_4$ is the second time derivative of the GW strain, it is required to integrate it twice in time to obtain $h_{lm}$. For the time integration and further GW signal analysis, we used a publicly available Python module, \codeword{Kuibit} \citep{kuibit_2021}, and computed $h_+$ and $h_x$  with Eq. \ref{eq:strain} via the FFI method described in \citep{Reisswig_2011}:
\begin{equation}
h_+^{lm}(r,t) - ih_\times^{lm}(r,t)  = 
\int_{-\infty}^t {du \int_{-\infty}^{u}{dv~\psi_4^{lm}(r,v)}}
\label{eq:strain}
\end{equation}
For all of our GW analysis, the complex combination of the extracted GW amplitude polarizations, $h=h_+ - ih_\times$, is used, and only the dominant mode, i.e. $l=m=2$, is considered. Note that the $h_{22}$ mode we extracted from the simulation is the amplitude of the (2,2) mode without the spherical harmonics term. Therefore it does not take into account the effect of the viewing angle. We also plot the waveform as a function of retarded time (Eq. \ref{eq:tret}), where the areal radius is $R = \sqrt{A(r)/4\pi}$, and $A(r)$ is the surface of the sphere of coordinate radius $(r)$:
\begin{equation}
t_{ret}\ =\ t\ - R - 2M_{0}\ log(\frac{R}{2M_{0}}-1).
\label{eq:tret}
\end{equation}
The instantaneous frequency of GW is computed as 
\begin{equation}
f_{GW}\ =\ \frac{1}{2\pi} \frac{d\phi}{dt},
\label{eq:instfreq}
\end{equation}
where $\phi = arctan(h_\times / h_+)$ is the phase of the GW signal. The frequency at merger is $f_{\rm merger}=f_{GW}(t=0)$.

\begin{figure*} 
\centering
\subfloat{{\includegraphics[width=.45\linewidth]{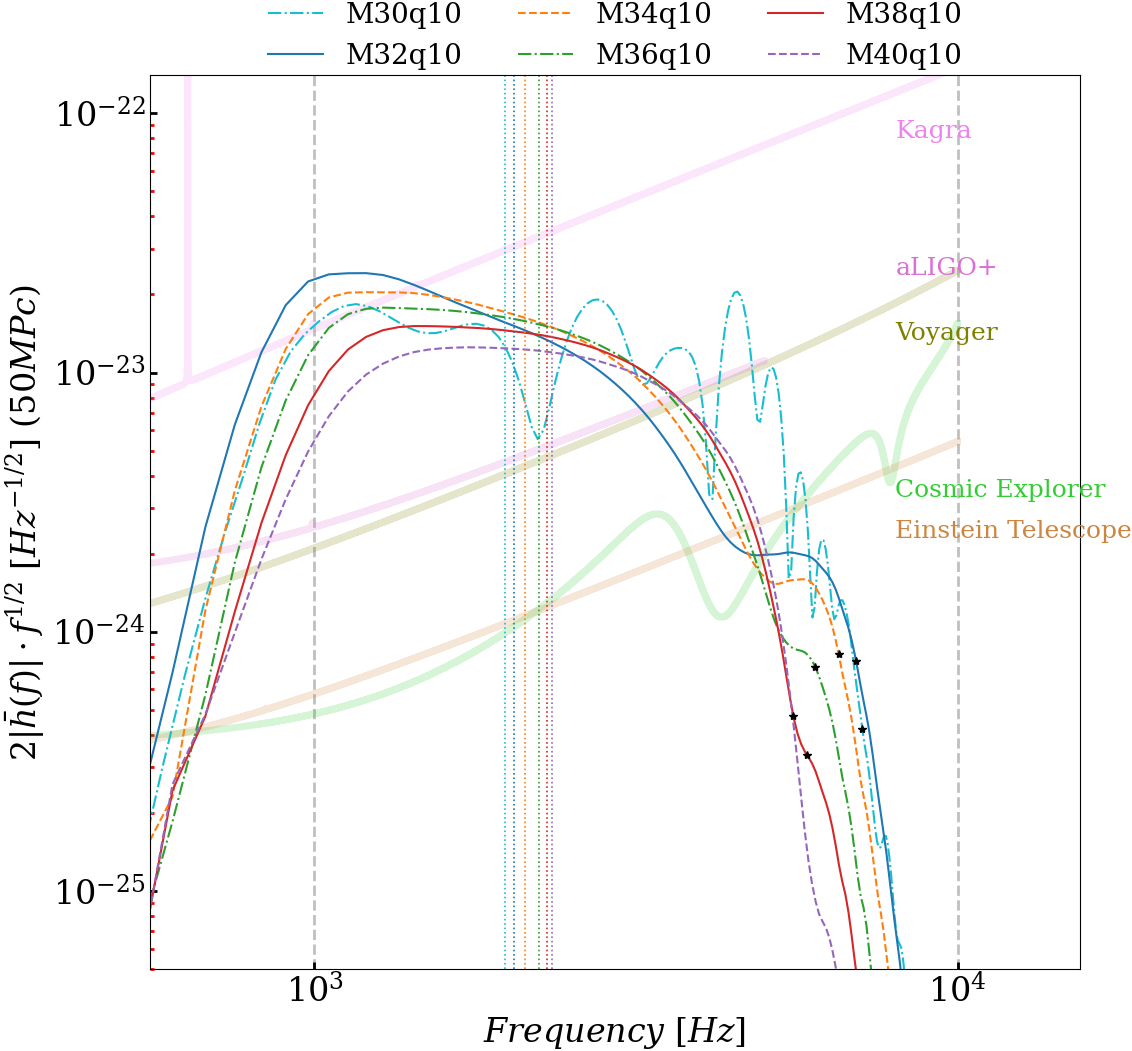}}}
\qquad
\subfloat{{\includegraphics[width=.45\linewidth]{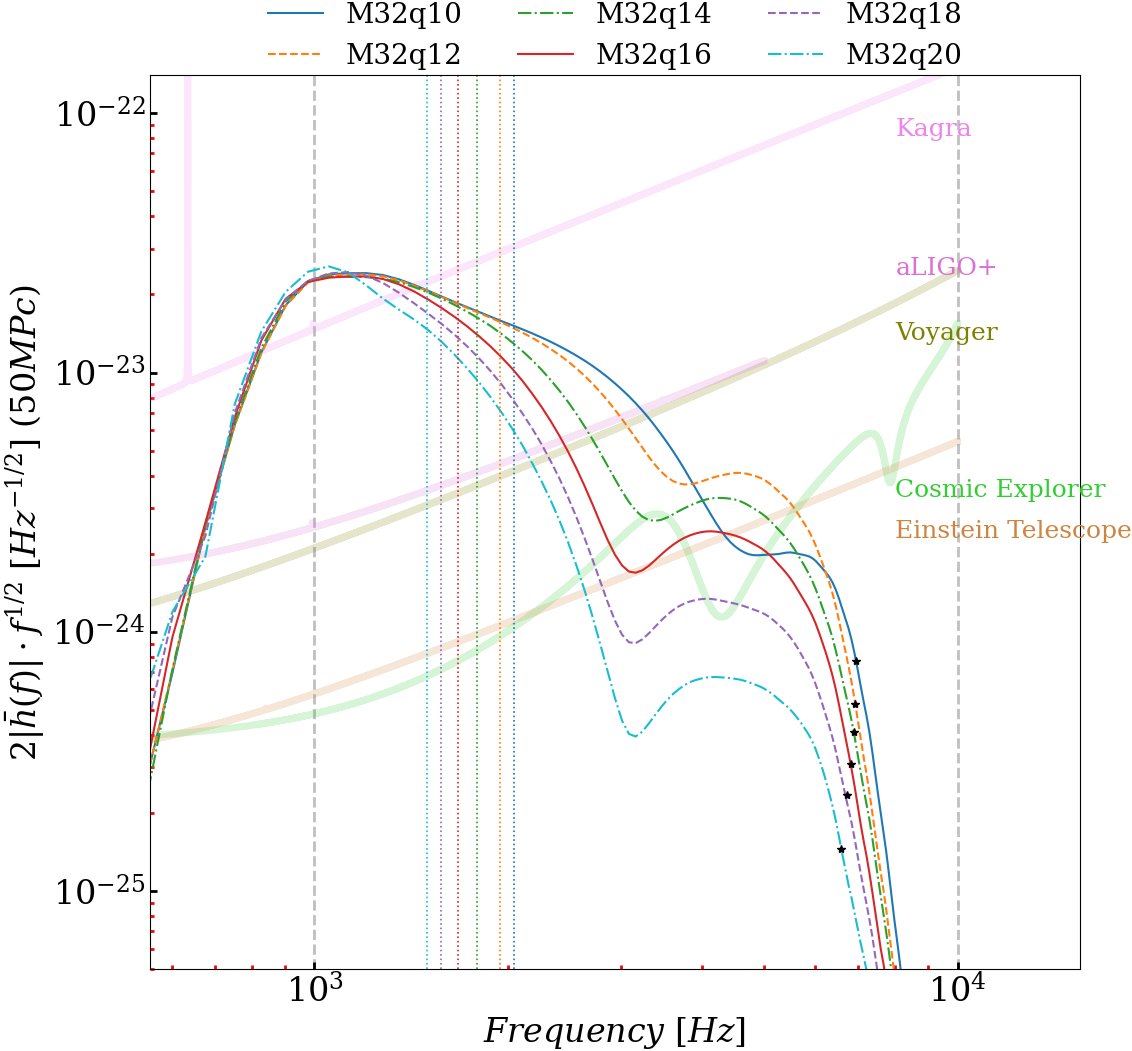}}}
\caption{\label{fig:ASD} Amplitude spectral densities (ASD) of the GW signals placed at a distance of $50Mpc$. The sensitivity curves of the current and the future-planned detectors (\href{https://granite.phys.s.u-tokyo.ac.jp/svn/LCGT/trunk/sensitivity/spectrum/BW2009_VRSED.dat}{KAGRA}, \href{https://dcc.ligo.org/public/0149/T1800042/005/AplusDesign.txt}{aLIGO+}, \href{https://dcc.ligo.org/LIGO-T1500293-v11/public}{Voyager}, \href{https://cosmicexplorer.org/data/CE1_strain.txt}{Cosmic Explorer} and \href{https://apps.et-gw.eu/tds/?content=3&r=14065}{Einstein Telescope}) are shown as shaded thick lines and they are the ones included in the public version of the \codeword{Kuibit} Python library~\citep{kuibit_2021}. Vertical dotted lines correspond to the merger frequencies estimated from the instantaneous frequency at the merger. The quasi-normal-mode frequencies of the final black holes are marked with black star markers.}
\end{figure*}

The GW strain (upper) and the phase velocity (bottom) for equal (left) and unequal (right) mass models are shown in Figure~\ref{fig:GWFreq}. Gravitational waveforms include the last part of inspiral, approximately three orbits before merger, and the coalescence and ringdown stages for all models. Models with larger total mass show, as expected, higher GW amplitudes at merger, while models with higher mass ratios have smaller amplitudes.

Because of the prompt formation of a BH, all GW signals go to zero less than $\approx 2$ ms after the merger. The frequency evolution for all models is shown in the bottom panels. Before the merger stage, the frequencies are nearly identical for all models and increase monotonically with GW amplitude in time. Except for high mass ratio models, the frequency increases after merger due to increased compactness and faster rotation of the remnant, as explained in \citet{Endrizzi_2016} and \citet{Kastaun_2016}. They behave differently for the latter in that, while model \model{M32q20} oscillates around $1\ kHz$, the frequencies of the other models rapidly increase shortly before $1\ ms$.

We then compute the amplitude spectral density (ASD) as $2|\bar{h}(f)|f^{1/2}$, where $\bar{h}(f)=\sqrt{\frac{|\bar{h}_+(f)|^2 + |\bar{h}_\times(f)|^2}{2}}$, where $\bar{h}_+$ and $\bar{h}_\times$ are the Fourier transforms of $h^{22}_+$ and $h^{22}_\times$ computed from the beginning of the simulations up to $10ms$ after merger. Figure \ref{fig:ASD} shows the amplitude spectral density of the GW signals for equal (left panel) and unequal (right panel) models placed at a distance of $50\ Mpc$ and the sensitivity curves of current (aLIGO+, KAGRA) and future planned (Einstein Telescope, Voyager, Cosmic Explorer) detectors. The initial frequency peak at around $1000\ Hz$ is equal to double of the initial orbital frequency of each model, and it just indicates the beginning of the inspiral stages in our simulations. From Figure \ref{fig:ASD} we can see that all systems can be observed in their merger phase by all detectors but KAGRA (which would be able to see only the earlier part of the inspiral, not simulated here).

Besides, we observe plateaus between $4000-7000\ Hz$ for all models, but \model{M38q10} and \model{M40q10} which are those that form a BH earlier. For the unequal-mass models, the frequency of plateaus changes with the mass ratio and is more evident for larger mass ratios. Moreover, we realise that frequencies of plateaus show a turning point. While a more noticeable plateau is observed in models with higher mass ratios, the clarity decreases as we move towards models with equal mass. Moreover, plateaus are observed at higher frequencies and higher ASD values as the mass ratios decrease, while the ASD value decreases after $q=1.0-1.2$. This suggests that there is a turning point in this mass ratio range.
In \citet{Kiuchi_2010} it is suggested that plateaus at high frequencies are related to both the formations and evolution of black holes and their surrounding disks. Therefore, we compare the ASD of model \model{M30q10}, which shows delayed collapse into a BH surrounded by torus after a short-lived HMNS phase, and prompt collapse models with torus (e.g. \model{M32q10}) and without torus (e.g. \model{M34q10, M36q10}). We realise that those plateaus are only related to the case of prompt collapse models regardless of having a torus or not. 

We calculated the quasi-normal-mode frequencies of the corresponding BH mass and spin using Eq. \ref{eq:fqnm} from \citet{Nakamura_1987}.
\begin{equation}
f_{qnm} \approx
10.8 \left(\frac{M}{3M_\odot}\right)^{-1} [1-0.63(1-\chi)^{-0.3}] kHz,
\label{eq:fqnm}
\end{equation}
According to \citet{Kiuchi_2010}, the peak frequency and width of the plateau show a correlation with the disk mass. Thus, these plateaus provide us with information about the formation and evolution of matter around the central object. If the frequency increases, we observe a decrease in the amplitude spectral density, as expected, due to the QNM ring-down of the formed BH. In Fig. \ref{fig:ASD}, the black star marks show the QNM ring-down frequencies for each model.

\begin{table*}
\centering
\caption{\label{tab:results}
We report the values of numerical results from the standard resolution (SR) simulations and the value inside brackets (third row) shows the error on the last digit as the absolute semi-difference between low resolution (LR) and high resolution (HR) simulations, except for the disk and ejecta masses where we provide the relative errors. Details on the estimates of such errors can be found in the Appendix \ref{appendix:A}. The second column shows the instantaneous frequency at merger, \fmerger, corresponding to the maximum GW amplitude. The third column is the elapsed time, $t_{BH}$, from the merger to the formation of BH. $M_{rem}$ is the mass of the remnant. If the remnant is a BH, it is a quasi-local measure mass, otherwise, it is the ADM mass of the remnant NS. The fifth to last columns represent, respectively, the dimensionless spin of the BH, $\chi_{BH}$, the disk mass, $M_{\rm disk}$, the amount of ejected matter, $M_{ejec}$ and total radiated energy in GW, $E_{GW}$.}
\begin{tabular}{lcccccccr}
\hline
{Model}	&$f_{\rm merger}$	&$t_{BH}$	&$M_{\rm rem}$	            &$\chi_{\rm BH}$	&$M_{\rm disk}$              &$M_{\rm ejec}$                 &$E_{\rm GW}$	\\
  		& $(kHz)$  		&$(ms)$  	&$(M_\odot)$	        &$(J/M^2)$ 		&$(10^{-3}~M_\odot)$     &$(10^{-3}~M_\odot)$        &$(M_\odot c^2)$ \\ 
  		&$(2)$          &$(1)$      &$(1)$                  &$(6)$          &$(36\%)$                &$(96\%)$                   &$(2)$\\
 \hline
M28q10	&1.87           &-  	    &2.133                  &-	            &660                     &2                       &0.065  \\
\hline
M30q10	&1.98	        &2.61	    &2.613	                &0.727	        &20                      &7                       &0.056  \\
M30q11	&2.01	        &-	        &2.225          	    &-	            &800                     &20                       &0.063  \\
\hline
M32q10	&2.05	        &0.72	    &2.823	                &0.790	        &0.11                    &0.03	                    &0.029  \\
M32q11	&2.01	        &0.75	    &2.818	                &0.789	        &3                       &0.8                       &0.030	\\
M32q12	&1.94	        &0.79	    &2.802	                &0.779	        &17                      &6                       &0.029	\\
M32q13	&1.86	        &0.84	    &2.777	                &0.766	        &40                      &12                       &0.026	\\
M32q14	&1.79	        &0.87	    &2.760	                &0.756	        &60                      &11                       &0.023	\\
M32q16	&1.67	        &0.86	    &2.722	                &0.726	        &95                      &9                       &0.019	\\
M32q18	&1.57	        &0.83	    &2.700	                &0.698	        &108                     &11                       &0.015	\\
M32q20	&1.50	        &0.79	    &2.680	                &0.663	        &113                     &17                       &0.013	\\
\hline
M34q10	&2.13	        &0.57	    &2.973	                &0.779	        &-                       &-                          &0.034	\\
\hline
M36q10	&2.24	        &0.47	    &3.118	                &0.768	        &-                       &-                          &0.042	\\
M36q11	&2.20	        &0.48	    &3.118	                &0.769	        &-                       &-                          &0.041	\\
\hline
M38q10	&2.30	        &0.42	    &3.258	                &0.754	        &-                       &-                          &0.051	\\
M38q11	&2.27	        &0.42	    &3.258	                &0.755	        &-                       &-                          &0.050	\\
\hline
M40q10	&2.34	        &0.37	    &3.393	                &0.741	        &-                       &-                          &0.061	\\
M40q11	&2.33	        &0.38	    &3.394	                &0.741	        &-                       &-                          &0.059	\\
\hline
\end{tabular}
\end{table*}

\subsection{Remnant Properties}
\label{sec:remnant}

We consider the matter whose time-component of the four-velocity is $u_t < -1$ as ejected matter, and we indicate its mass as $M_{ejec}$. We also define the disk mass as the baryonic mass of regions with rest mass density above the artificial atmosphere, $\rho_{atm} = 6.2\times10^6 g/cm^3$, and that is not ejected (i.e., we exclude $M_{\rm ejec}$ from the disk mass). Besides, to consider the effects of the artificial atmosphere's density to disk mass and ejected matter, we simulated a model of $\model{M32q10}$ with three different atmosphere densities. Details are given in Appendix \ref{appendix:A}.

All simulations run in this study result in the rapid collapse into a BH, except \model{M2810, M30q11} (hyper-massive NS) and \model{M30q10} (delayed collapse into a BH). In Table \ref{tab:results} we report the formation time of the BH after the merger. As seen from the values of $t_{BH}$, all promptly collapse models form a BH before $0.9\ ms$. The results suggest that for massive binary systems, the rapid formation of a BH occurs sooner because there is not enough time for the redistribution of angular momentum to support the remnant, resulting in an earlier collapse into a BH. 
As suggested in \citet{Shibata_2003} and \citet{Shibata_2006}, nearly all the matter tends to fall into the BH in the prompt collapse scenario. Moreover, it is realised that the elapsed time for the formation of a BH changes in a parabolic way with the increase of the mass ratio. When the mass ratio is between $q=1.0-1.4$, the required time to form a BH increases, but after $q=1.4$, it starts to decrease.

We report the final mass and dimensionless spins of BH found by \codeword{AHFinder} after the merger in Table \ref{tab:results}. The final black hole swallows $94-99\%$ of the initial binary mass. Similarly, we estimate that approximately $70-80\%$ of the initial total angular momentum is transferred to the final BH with dimensionless spin between $0.66-0.79$ as also suggested in \citet{Kiuchi_2009}. We also estimated that the energy carried away by gravitational radiation should be between $0.01-0.07 M_\odot c^2$. 

\begin{figure*}
\includegraphics[width=1.\textwidth]{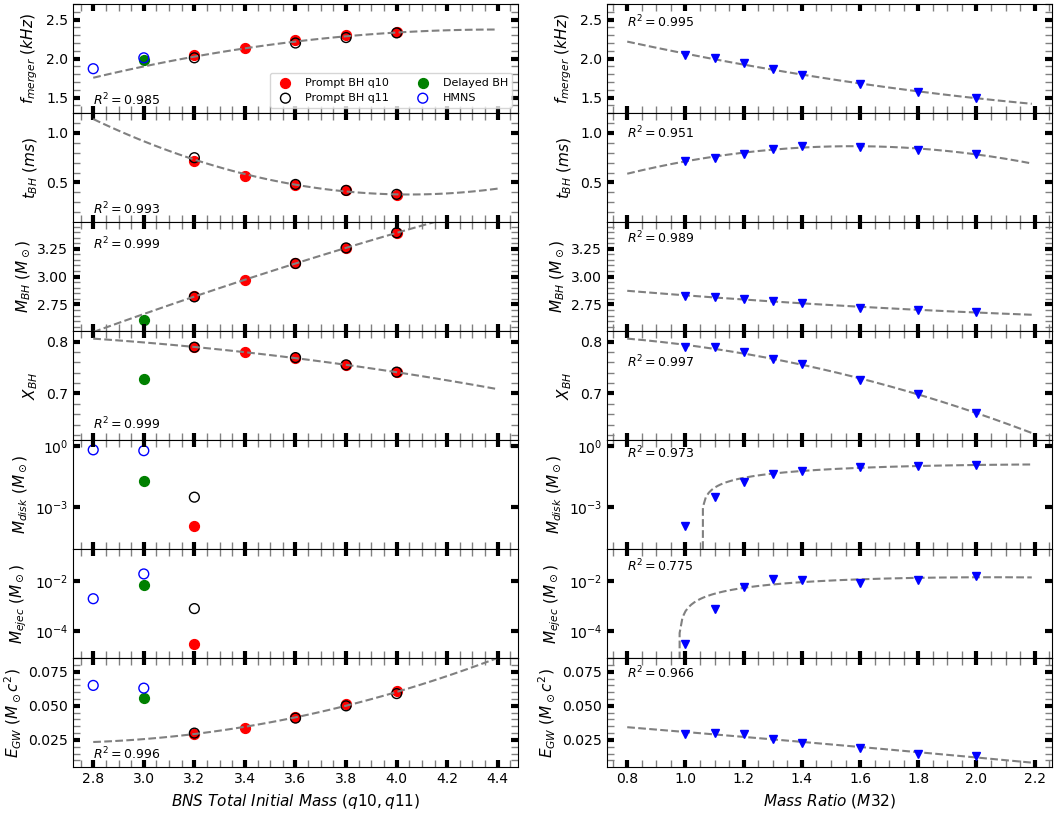}
\caption{\label{fig:scalars}
Dependence of the numerical results given in Table \ref{tab:results} from total initial mass (left panels) and mass ratios (right panel). Dashed lines on both plots are fits calculated by considering only results from promptly collapsed models.
}
\end{figure*}

Figure~\ref{fig:scalars} shows the dependence of $f_{\rm merger}$, $t_{\rm BH}$, $M_{\rm BH}$, $E_{\rm GW}$, $M_{\rm disk}$ and $M_{\rm eject}$  on the different initial total mass of the system and mass ratios for the eighteen simulations performed in this study. As can be seen in Figure~\ref{fig:scalars}, the relationships between some of the parameters are clearly visible.

\begin{equation}
\begin{aligned}
f_{\rm merger}  & =  -0.242\ M^2_b + 2.132\  M_b - 2.316  \,(1.5\%) \\ 
                & =   0.166\  q^2 - 1.070\ q + 2.968 \,(1.5\%)   \\  \\
t_{\rm BH}      & =   0.491\ M^2_b - 3.971\  M_b + 8.413 \,(1.1\%)   \\ 
                & =  -0.469\  q^2 + 1.480\  q - 0.302 \,(1.3\%)   \\  \\
M_{\rm BH}      & =  -0.069\ M^2_b + 1.212\  M_b -0.351  \,(0.1\%)  \\ 
                & =   0.041\  q^2 - 0.279\  q + 3.069  \,(0.6\%)  \\  \\
\chi_{\rm BH}  & =  -0.017\ M^2_b + 0.064\  M_b + 0.762  \,(0.1\%)  \\  
                & =  -0.059\  q^2 + 0.047\  q + 0.806  \,(0.2\%)  \\  \\
E_{\rm GW}      & =   0.023\ M^2_b - 0.127\  M_b + 0.201  \,(0.1\%)  \\
                & =  -0.001\  q^2 - 0.017\  q + 0.049   \,(0.1\%) \\ \\ 
M_{\rm disk}    & =  -0.079\  q^2 + 0.366\  q - 0.298   \,(0.8\%) \\  \\
M_{\rm eject}   & =  -0.013\  q^2 + 0.052\  q - 0.039  \,(0.3\%) \\   \\             
\end{aligned}
\label{eq:st1}
\end{equation}

We plot in Fig.~\ref{fig:scalars} E$_{\rm GW}$, M$_{\rm eject}$, M$_{\rm disk}$, $\chi_{\rm BH}$, M$_{\rm BH}$, t$_{\rm BH}$, and f$_{\rm merger}$ in function of the initial total baryonic mass of the BNS $(M_b)$ and mass ratio $(q)$ (Table~\ref{tab:results}). The dashed black curve is fit to a second-order polynomial using statistical analysis given by Equation~\ref{eq:st1}. The uncertainties in the equations are expressed as percentages in parentheses. We believe these relations can be used as a prediction for future simulations using similar equations of state.

In Fig. \ref{fig:scalars}, we show how the values extracted from our simulations change with initial binary neutron star system mass (left panels) and mass ratio (right panels). From the left panels in the figure, it can be seen that the dimensionless spin (mass) of the BH decreases (increases) with increasing BNS mass. When increasing the BNS mass, also the disk mass and the amount of ejected matter from the system decrease. Energy radiated via GW is also increasing for higher mass binary neutron star systems. From these plots, we can estimate that, in the prompt collapse case, the BH swallows $98-99\%$ of the initial mass of the BNS systems and that the remaining $1-2\%$ of the initial mass is radiated away from the system as $E_{GW}$. On the other hand, for systems with different mass ratios (right panel of Fig. \ref{fig:scalars}) the final mass of the BH decreases with the increase of mass ratio while the BH spin decreases. Similarly, we observe more massive disks and a higher amount of ejected matter, which are also consistent with the decrease in $E_{GW}$ for higher $q$.

\subsection{Dynamics of Systems During Evolution}
\label{sec:grhd}

\begin{figure*}
\includegraphics[width=0.95\linewidth]{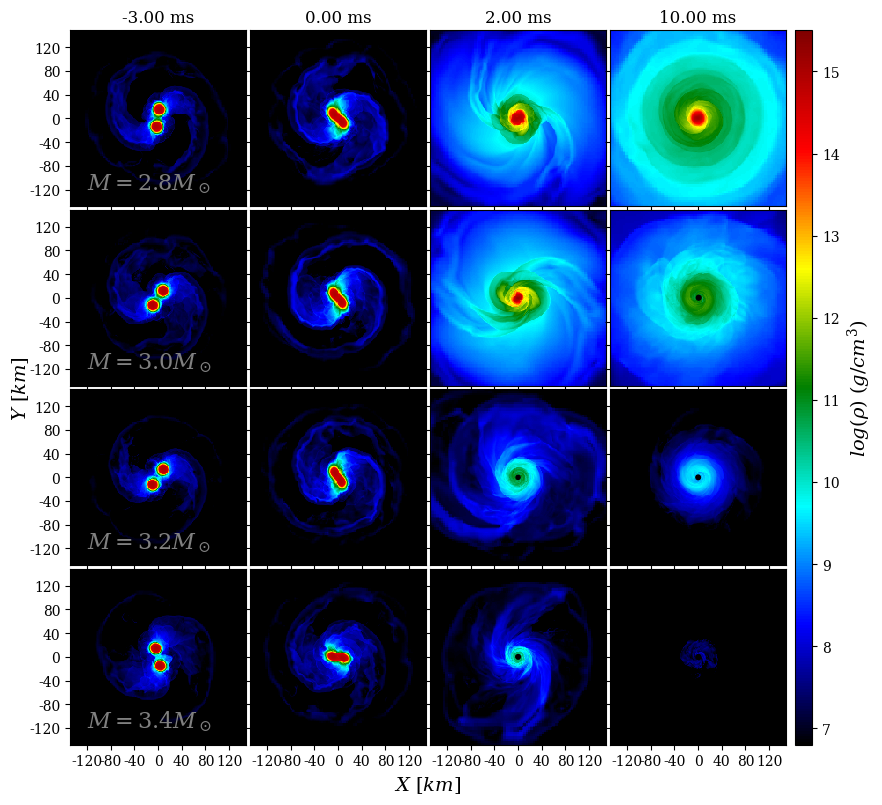}
\includegraphics[width=0.95\linewidth]{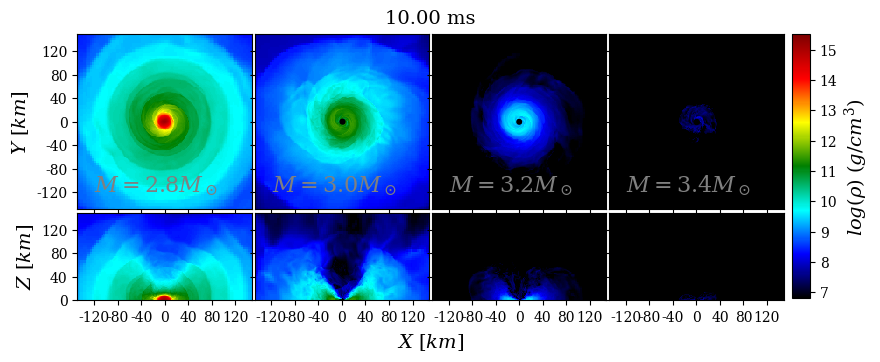}
\caption{\label{fig:equalRhoXYZ} Snapshots of the rest mass density evolution for the equal mass models. Upper panels: the rest-mass density is plotted on the equatorial plane $3\ ms$ before the merger, at merger $t=0$, $2$ and $10\ ms$ after the merger. Each row represents a different model: top to bottom, \model{M28q10, M30q10, M32q10} and \model{M34q10}. Bottom panels: comparison of the rest mass density snapshots in the equatorial and meridional planes at the end of the simulations ($t=10.00\ ms)$.}
\end{figure*}

\begin{figure*}
\includegraphics[width=0.95\textwidth]{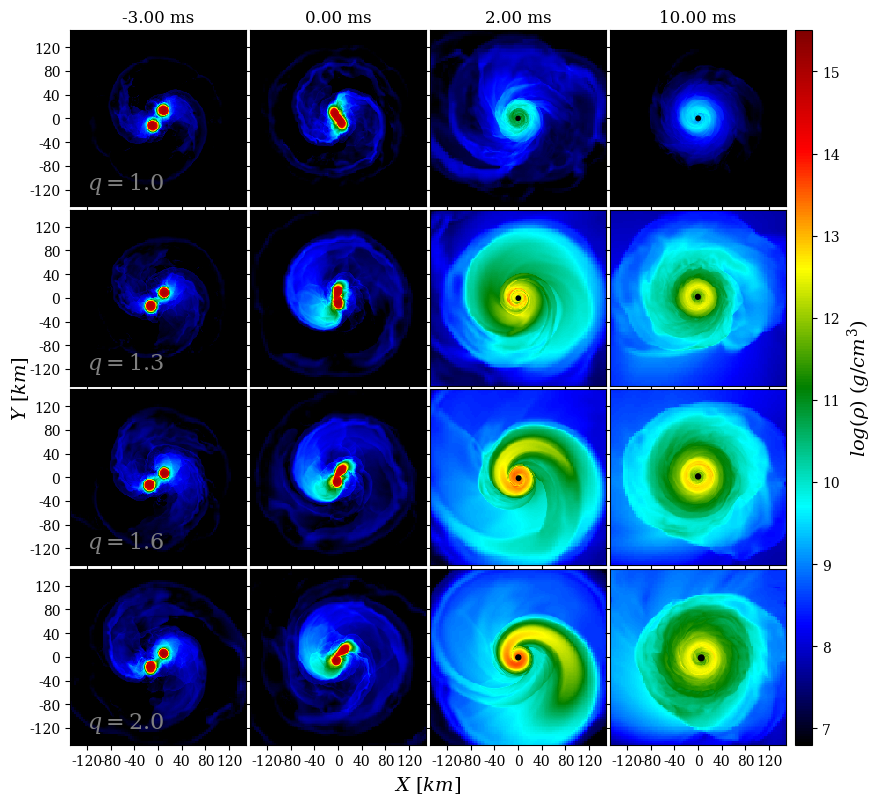}
\includegraphics[width=0.95\textwidth]{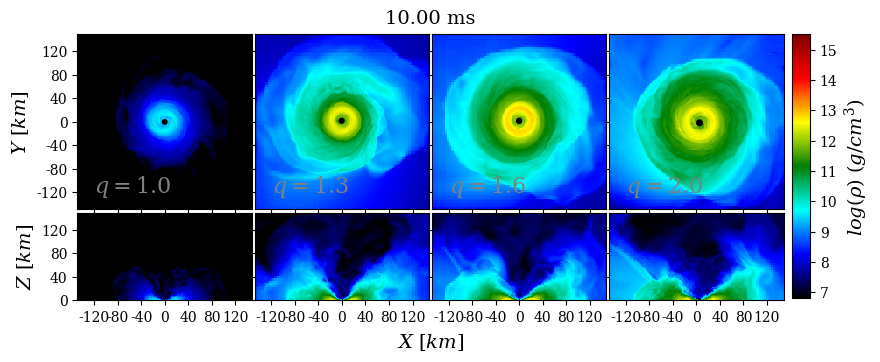}
\caption{\label{fig:unequalRhoXYZ} Snapshots of the rest mass density evolution for the unequal mass models. Upper panels: the rest-mass density is plotted on the equatorial plane $3\ ms$ before the merger, at merger $t=0$, $2$ and $10\ ms$ after the merger. Each row represents a different model: top to bottom, \model{M32q10, M32q13, M32q16} and \model{M32q20}. Bottom panels: comparison of the rest mass density snapshots in the equatorial and meridional planes at the end of the simulation ($t=10.00\ ms$).}
\end{figure*}

In Fig. \ref{fig:equalRhoXYZ}, we present the evolution of the rest mass density for the equal mass BNS models (\model{M28q10, M30q10, M32q10} and \model{M34q10}) in the equatorial plane (upper panel) at times $-3,\ 0,\ 2,\ 10\ ms$. The columns indicate times, while the rows refer to the models. All models are inspiralling and losing matter from the tidal tails due to tidal emission $3\ ms$ before the merger (first column in Fig. \ref{fig:equalRhoXYZ}). At $t=0.00\ ms$ (second column in Fig. \ref{fig:equalRhoXYZ}), the cores of the two neutron stars merge. Due to the high orbital speed (i.e., high angular momentum), matter is released in a spiral flow. After $2\ ms$ (third column in Fig. \ref{fig:equalRhoXYZ}), we show that \model{M32q10} and \model{M34q10} collapse into a BH which rapidly accretes the surrounding matter. However, model \model{M30q10} shows the formation of a short-lived HMNS right after the merger and the formation of a BH $2.6\ ms$ after the merger. Similarly, the remnant of \model{M28q10} is a high-mass neutron star with a thick torus. The comparison of the four models in the figure shows that the disk structure changes with the initial binary mass. The mass of the disk is crucially decreasing for high-mass BNS systems. The bottom panel of Fig. \ref{fig:equalRhoXYZ} shows the four models on the meridional plane. 

Besides, in Fig. \ref{fig:unequalRhoXYZ}, we present the evolution of the rest mass density for the BNS system with an initial total baryonic mass of $3.2$\Msun and different mass ratios in the equatorial plane (upper panel) at times $-3,\ 0,\ 2,\ 10\ ms$. The rows indicate the models \model{M32q10, M32q13, M32q16} and \model{M32q20} from up to bottom. The figure shows that models with higher mass ratios produce more massive and extended tidal tails and more matter is released during the inspiral phase. The increment in the amount of released matter is due to the smaller compactness of one of the neutron stars in the system, so that the matter from it can be easily unbound. 
In the case of model \model{M32q20}, the less massive NS is swallowed by the more compact one during the coalescence. Due to the less compact NS's high orbital speed, there is a one-armed spiral structure around the BH after the merger (third column). During this stage, the matter ejected during the merger becomes unbound due to the high angular momentum. At the end of the simulation (fourth column) we can also see how higher mass ratios produce more massive and more extended accretion disks (see also \citep{Rezzolla2010}).

The study of \citet{Bernuzzi_2020} reports that in the neutron star systems with mass ratios $q=1.5-2$, the massive primary NS tidally disrupts the companion, and the matter of the latter accretes onto the former one. These processes can make the remnant unstable and rapidly collapse into a BH. This process is named accretion-induced prompt collapse in \citet{Bernuzzi_2020}. Our model, \model{M32q20}, shows indeed a rapid accretion-induced collapse to BH.

\begin{table}
\centering
\caption{\label{tab:fitcomp} Relative differences between the fits and our numerical data for the ejected matter and disk mass. $\Delta M = M_{sim}-M_{fit}$ where $M_{sim}$ are the values given in Table \ref{tab:results} as $M_{\rm disk}$ and $M_{\rm eject}$ while $M_{fit}$ are the values computed using Eq. \ref{eq:ejecta_fit_eq} and \ref{eq:disk_fit_eq}.}
\begin{tabular}{lcr}
\hline
{Model}				    &$\Delta M / M_{\rm  eject}$	    &$\Delta M / M_{\rm disk}$ \\
\hline
\model{M28q10}			&$-0.34$		            &$-0.61$\\
\hline
\model{M30q10}			&$0.46$		                &$-49.55$\\
\model{M30q11}			&$0.81$		                &$-0.48$\\
\hline
\model{M32q10}			&$-161.34$                  &$-9057.10$ \\
\model{M32q11}			&$-4.66$		            &$-395.03$\\
\model{M32q12}			&$0.31$		                &$-79.98$\\
\model{M32q13}			&$0.68$		                &$-37.69$\\
\model{M32q14}			&$0.68$		                &$-27.12$\\
\model{M32q16}			&$0.63$		                &$-18.26$\\
\model{M32q18}			&$0.75$		                &$-15.25$\\
\model{M32q20}			&$0.84$		                &$-12.18$ \\
\hline
\end{tabular}
\end{table}

\subsection{Comparison of Our Results with Other Works}
\label{sec:comparison}

In the study of \citet{Kolsch_2022}, the authors used the BAM code to simulate models with three different resolutions: $dx=0.13,~ 0.16,~ 0.21$, which are closer to the resolutions we used in this study. The models employing the SLy EOS and a gravitational mass of $M=2.90 M_\odot$ at different mass ratios ($q=1.0,~ 1.25$ and $1.50$) evolved in \citet{Kolsch_2022} are close to our models \model{M32q10} $(M=2.88,~ q=1.0)$, \model{M32q13} $(M=2.88,~ q=1.26)$, and \model{M32q16} $(M=2.88,~ q=1.52)$. We compared the time for the formation of the BH after the merger, the BH mass, the BH dimensionless spin, and disk mass. For the models very similar to the \model{M32q10}, \model{M32q13} and \model{M32q16}, they reported respectively, $M_{BH} = 2.831, \sim 2.784, \sim 2.741\ M_\odot$, $\chi = 0.768, \sim 0.717, \sim 0.657$ and $M_{\rm disk} = 5.6 \times 10^{-4}, \sim 5.346 \times 10^{-2}, \sim 1.145 \times 10^{-1}\ M_\odot$. Besides the differences due to using different resolutions and numerical relativity codes, the values they reported are close to the ones we reported in Table \ref{tab:results} in this study.

We also compared our results with current fits available in the literature. For this reason, we used the statistical studies on dynamical ejecta and disk mass in \citet{Nedora_2021} to check whether our results are in agreement with their fits. We use in particular the fits for the dynamical ejecta and disk mass listed below (Eq. \ref{eq:ejecta_fit_eq} and \ref{eq:disk_fit_eq}):
\begin{equation}
P^2_2(q, \widetilde{\Lambda})= 
b_0 + b_1 q + b_2 \widetilde{\Lambda} + b_3 q^2 + b_4 q \widetilde{\Lambda} + b_5 \widetilde{\Lambda}^2 \,;
\label{eq:ejecta_fit_eq}
\end{equation}
\begin{equation}
log_{10}(P^2_2(q, \widetilde{\Lambda}))= 
log_{10} (b_0 + b_1 q + b_2 \widetilde{\Lambda} + b_3 q^2 + b_4 q \widetilde{\Lambda} + b_5 \widetilde{\Lambda}^2).
\label{eq:disk_fit_eq}
\end{equation}
We use the coefficients recommended in \citet{Nedora_2021}: $b_0 = -1.32$, $b_1 = -0.382$, $b_2 = -4.47 \times 10^{-3}$, $b_3 = -0.339$, $b_4 = 3.21 \times 10^{-3}$, $b_5 = 4.31 \times 10^{-7}$ for the dynamical ejecta, and $b_0 = -1.85$, $b_1 = 2.59$, $b_2 = 7.07 \times 10^{-3}$, $b_3 = -0.733$, $b_4 = -8.08 \times 10^{-4}$, $b_5 = 2.57 \times 10^{-7}$ for the disk mass.

The results of these formulae are reported in Table \ref{tab:fitcomp}. Except for model \model{M32q10}, the amount of dynamical ejecta we estimated in our simulations agrees with the fits calculated using Eq. \ref{eq:ejecta_fit_eq}. Furthermore, the value of ejected matter from model \model{M32q11} is consistent with the trend. When we compare the disk masses estimated in this study to those estimated with the fit of \citet{Nedora_2021}, we see that our computed values are always less than the disk mass obtained from the fit. However, such differences may also be related to the various methods and times in which the disk mass was estimated. Moreover, \citet{Camilletti_2022} reported that among the four EOSs (BLh, DD2, SFHo, SLy), the SLy EOS used in our simulations provides the lightest disk mass around a central BH. This could also explain the discrepancy between our numerical results and the fit estimates.

According to simulations run in \citet{Radice_2018} it is proposed that BNS systems with $\widetilde{\Lambda}<450$ inevitably promptly collapse into a BH and form a surrounding disk $<0.02$\Msun and that systems with $\widetilde{\Lambda}<400$ could not be compatible with a lower limit of ejected matter inferred from UV/optical/infrared observations of AT2017gfo, which is the optical counterpart to GW170817. However, it is also stated in \citet{Radice_2018} and references therein that mass ratios higher than $1.25$ might also form disks with masses up to $0.1 M_\odot$. According to the outcomes of the models consistent with the chirp mass of GW170817 in this study (\model{M30q10, M30q11, M32q18}), models \model{M30q11} and \model{M32q18} with, respectively, $\widetilde{\Lambda}=335, 347$ may eject at least $0.05$\Msun. 
Model \model{M30q11} may unbound at least $0.05$\Msun matter if the disk has an ejection efficiency around $5\%$. For our promptly collapse models, \model{M32q18}, this contribution should be at least $35\%$ if we consider the high-spin assumption for GW170817. Therefore, as stated in \citet{Kiuchi_2019}, we also show that the unequal-mass systems with $\widetilde{\Lambda}<400$, even if they collapse into a BH promptly, provide both the lower limit of ejected mass, $0.05$\Msun, and disk mass, $0.08$\Msun, reported in \citet{Perego_2017}. Thus, according to the simulations we run in this study we may confirm that the lower bound on $\widetilde{\Lambda}$ might depend on the mass ratio as suggested in \citet{Radice_2018}.


\section{Conclusions}
\label{sec:conclusions}
We performed a set of equal and unequal high-mass BNS system simulations with seven-segment piecewise polytropic SLy EOS in this study. We started the simulations with $3$ orbits before the merger and followed our systems up to $10\ ms$ after the merger. Except for models \model{M28q10, M30q10}, and \model{M30q11}, all models show a rapid collapse into a BH within $0.9\ ms$ and form a BH with mass range $M_{BH}=2.68-3.39M_\odot$ and dimensionless spin range $\chi_{BH}=0.66-0.79$. Our models also estimated the energy radiated in gravitational waves to be between $0.01-0.07M_\odot c^2$. 

We also investigated the relationship between $M_{BH},\ \chi_{BH}\ $ and $E_{GW}\ $, as well as the mass ratio and initial binary mass. We showed that, as expected, the disk mass and the amount of dynamically ejected matter either increase or decrease, respectively, with an increase in mass ratio and total binary mass. 

We reported our estimates for the disk mass and amount of ejected matter and compared them to the literature's current fits. Our models' estimates of mass ejection agree with the fit, but the amount of disk mass is less than the fit's estimate.

Moreover, we calculated gravitational waves and their spectra. We compared the amplitude spectral density of our models to the KAGRA, aLIGO+, Voyager, Einstein Telescope, and Cosmic Explorer sensitivity curves. We propose a possible feature in the amplitude spectral density that could indicate rapid collapse into a BH following the merger of two neutron stars, and that could be observed using future-planned detectors such as the Einstein Telescope and Cosmic Explorer. 

Additionally, from the evolution of the rest-mass density, we discussed the possible impact of initial binary mass and mass ratio on the geometry and mass of the accretion disk that will be formed around the post-merger NS or BH.


\section*{Acknowledgements}
The current study is part of the PhD thesis of KAC. This study was supported by the Scientific and Technological Research Council of Turkey (T\"{U}B\.{I}TAK 117F188 and 119F077). KAC thanks T\"{U}B\.{I}TAK for his Fellowship (2210-C and 2211-A). The work has been performed under Project HPC-EUROPA3 (INFRAIA-2016-1-730897), with the support of the EC Research Innovation Action under the H2020 Programme; in particular, KAC gratefully acknowledges the support of Bruno Giacomazzo, the University of Milano-Bicocca, Department of Physics "Giuseppe Occhialini" and the computer resources and technical support provided by HPC-CINECA. KY would like to acknowledge the contribution of COST (European Cooperation in Science and Technology)  Action CA15117 and CA16104.

\section*{Data Availability}

The initial datasets we used in these simulations are available in a repository and can be accessed via \href{https://zenodo.org/records/8382258}{10.5281/zenodo.8382258}.



\bibliographystyle{mnras}
\bibliography{bns2023_bib} 



\appendix

\section{Error Estimate, Mass Conservation and Effects of Artificial Atmosphere's Density}
\label{appendix:A}

In Table \ref{tab:results} we reported the values obtained from our simulations run with standard resolution (SR), i.e,  with an innermost grid resolution of $dx=0.16$. To estimate the uncertainty of our numerical results, we performed higher (HR) and lower resolution (LR) simulations for two models using respectively a resolution of $dx=0.12$ and $dx=0.20$ on the finest level. To compute our error, we used the method of absolute semi-differences given in Eq. \ref{eq:absdiff}
\begin{equation}
Err = \frac{|HR - LR|}{2},
\label{eq:absdiff}
\end{equation}
for models \model{M32q10} and \model{M32q11}. Estimated values for our errors are given in Table \ref{tab:uncert}. In the two models, the maximum value of error (last column of Table \ref{tab:uncert}) is considered as an error on related parameters. Note that due to the wide range of masses from $10^{-1}$ to $10^{-5}$, the errors on disk mass and amount of dynamical ejecta are given as uncertainties rather than absolute errors.

\begin{figure}
\includegraphics[width=.49\textwidth]{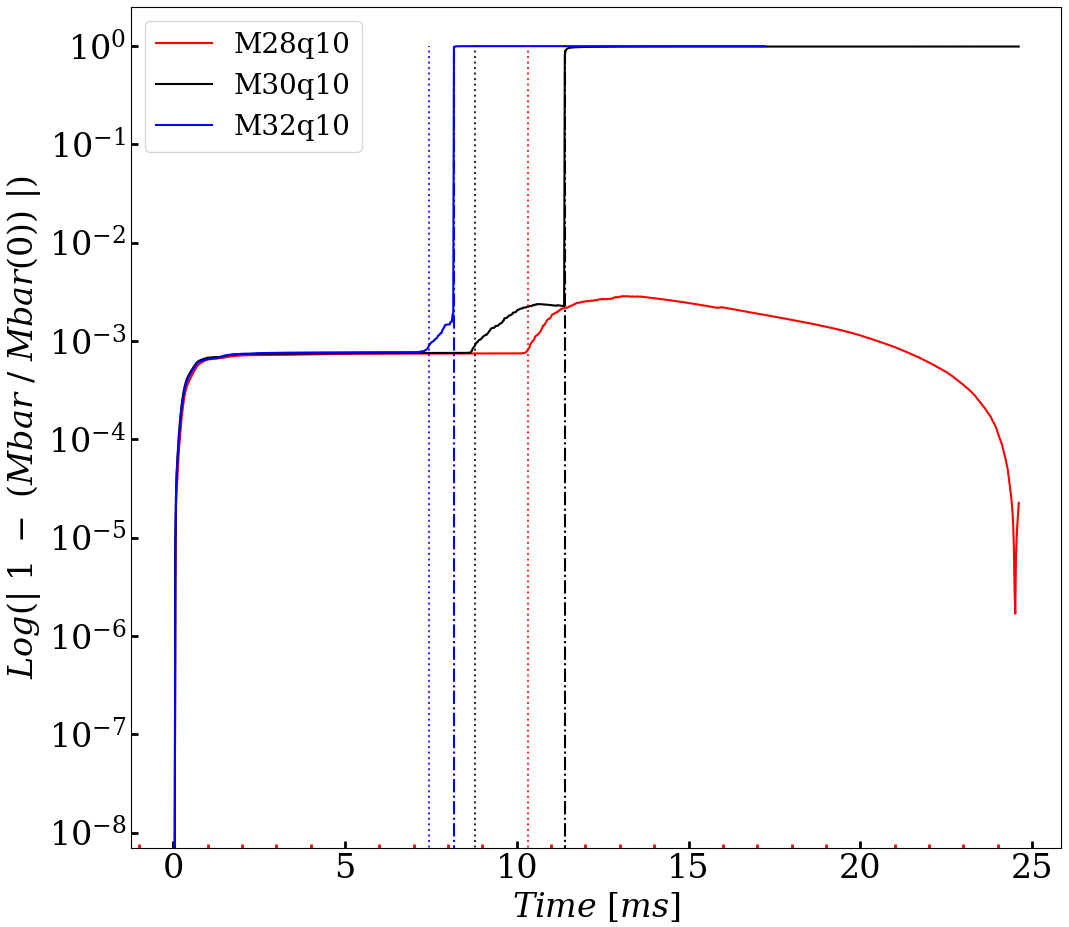}
\caption{\label{fig:mass_conservation}
Relative error on the baryon mass conservation for the model \codeword{M28q10}, \codeword{M30q10} and \codeword{M32q10}. The times of the merger and BH formation, if formed, are shown as vertical dotted and dot-dashed lines.
}
\end{figure}

We present the baryon mass conservation for models \codeword{M28q10}, \codeword{M30q10} and \codeword{M32q10} in Fig. \ref{fig:mass_conservation} and report the values for each model before merger and before BH formation in Table \ref{tab:mass_conservation}. The relative errors on baryon mass conservation before the merger and BH formation (or over the entire simulation for models \model{M28q10} and \model{M30q11}) are, respectively, below $1.4 \times 10^{-3}$ and $2.9 \times 10^{-3}$. In this case, the disk and ejecta masses used in Table~\ref{tab:results} for the models of M32q10 and M32q11 have values smaller than the baryonic mass conservation error. It should be taken into consideration that it is quite possible that the disk and ejecta mass values obtained from these models were affected by the baryonic mass conservation error.

To investigate the effect of artificial atmosphere density on disk mass and ejected matter, we ran simulations with three different artificial atmosphere densities for the model \model{M32q10} (see Table \ref{tab:atmos}). $(1\times 10^{-9})$, $(1\times 10^{-11})$ and $(5\times 10^{-12})$ for high, standard, and low artificial atmosphere density, respectively. We report that higher NS atmosphere density causes less disk mass and ejecta matter to be measured. On the other hand, if we use half of the standard atmosphere density, there is no difference in the amount of disk mass, but the amount of ejecta varies $1.6$ times. As a result, the disk mass should be considered accurate, whereas the ejecta error estimate is larger.

\begin{table}
\centering
\caption{\label{tab:uncert} Errors on each value given in Table \ref{tab:results} using \model{M32q10} and \model{M32q11}. Errors on $M_{\rm disk}$ and $M_{\rm ejec}$ are given as relative errors.}
\begin{tabular}{lccr}
\hline
{Quantity}		    &\model{M32q10}	    &\model{M32q11}     &Max Error    \\
\hline
$f_{\rm mer}$   		&$0.02$		        &$0.02$		        &$\pm 0.02$    \\
$t_{\rm BH}$ 			&$0.002$		    &$0.008$		    &$\pm 0.008$     \\
$M_{\rm BH}$ 	    	&$0.001$		    &$0.0001$	        &$\pm 0.001$     \\
$\chi_{\rm BH}$			&$0.006$		    &$0.003$		    &$\pm 0.006$    \\
$E_{\rm GW}$    		&$0.002$		    &$0.002$		    &$\pm 0.002$     \\
$M_{\rm disk}$         	&$36 (\%)$		    &$17 (\%)$		    &$36 (\%)$  \\
$M_{\rm ejec}$      	&$96 (\%)$		    &$15 (\%)$		    &$96 (\%)$   \\

\hline
\end{tabular}
\end{table}

\begin{table}
\centering
\caption{\label{tab:mass_conservation} We reported the maximum relative error on the baryon mass for the all models simulated in this paper. The symbol (*) indicates that the model does not have enough matter to collapse into a BH after the merger. Therefore, we reported the maximum baryon mass error during the whole simulation.}
\begin{tabular}{lcr}
\hline
{Model}		        &Max Value(\%)	    &Max Value(\%)   \\
                    &$t < t_{\rm mer}$	    &$t < t_{\rm BH}$ \\
\hline
\codeword{M28q10}	&0.08			&0.29*			    \\
\hline
\codeword{M30q10}	&0.09			&0.24				\\  
\codeword{M30q11}   &0.09			&0.26*				\\
\hline
\codeword{M32q10}	&0.09			&0.19				\\  
\codeword{M32q11}	&0.09			&0.25		 	\\
\codeword{M32q12}	&0.09			&0.24			\\
\codeword{M32q13}	&0.09			&0.25			\\
\codeword{M32q14}	&0.09			&0.17			   \\
\codeword{M32q16}	&0.08			&0.13				\\
\codeword{M32q18}	&0.07			&0.12			 	\\
\codeword{M32q20}	&0.07			&0.10				\\
\hline
\codeword{M34q10}	&0.09			&0.17		 	\\  
\hline
\codeword{M36q10}	&0.11			&0.22		 	\\  
\codeword{M36q11}	&0.11			&0.20			 	\\
\hline
\codeword{M38q10}	&0.13			&0.25				\\  
\codeword{M38q11}	&0.12			&0.22				\\
\hline
\codeword{M40q10}	&0.14			&0.29			\\  
\codeword{M40q11}	&0.14			&0.24		\\
\hline
\end{tabular}
\end{table}

\begin{table}
\centering
\caption{\label{tab:atmos} Disk and ejecta mass for simulation of the model \model{M32q10} with high, standard and low NS atmosphere density values.}
\begin{tabular}{lccr}
\hline
{Quantity}      &High	            &Standard           &Low  \\
                &$(1x10^{-9})$      &$(1x10^{-11})$     &$(5x10^{-12})$   \\
\hline
$M_{\rm disk}$      &$4.0\times 10^{-7}$      &$1.1\times 10^{-4}$      &$1.1\times 10^{-4}$	  \\
$M_{\rm ejec}$      &$4.6\times 10^{-8}$		&$3.3\times 10^{-5}$	    &$2.0\times 10^{-5}$	  \\
\hline
\end{tabular}
\end{table}


\bsp	
\label{lastpage}
\end{document}